# Precise probing spin wave mode frequencies in the vortex state of circular magnetic dots

A.A. Awad[1], K.Y. Guslienko[2,3], J.F. Sierra[1], G.N. Kakazei[4,*], V. Metlushko[5], and F. G. Aliev[1#]

[1] *Dpto. Física de la Materia Condensada, CIII, Universidad Autónoma de Madrid, 28049 Madrid, Spain*

[2] *Dpto. Fisica de Materiales, Universidad del Pais Vasco, 20018 Donostia-San Sebastian, Spain*

[3] *IKERBASQUE, the Basque Foundation for Science, 48011 Bilbao, Spain*

[4] *Departamento de Fisica da Faculdade de Ciencias, IFIMUP and IN – Institute of Nanoscience and Nanoterchnology, , Universidade do Porto, 4169-007 Porto, Portugal*

[5] *Dept. Electrical & Computer Engineering, University of Illinois at Chicago, Chicago, IL, 60607, USA*

We report on detailed broadband ferromagnetic resonance measurements of azimuthal and radial spin wave excitations in circular Permalloy dots in the vortex ground state. Dots with aspect ratio ($\beta$ =height over radius) varied from 0.03 to 0.1 were explored. We found that for $\beta$ exceeding approximately 0.05, variation of the spin wave eigenfrequencies with $\beta$ deviates from the predicted ~ $\sqrt{\beta}$ dependence. The frequency splitting of two lowest azimuthal modes was observed. The experimentally observed dependence of the frequency splitting on $\beta$ was reasonably well described by dynamic splitting model accounting the spin-waves and, vortex gyrotropic mode interaction.

[*] On leave from Institute of Magnetism National Academy of Sciences of Ukraine, 03142 Kiev, Ukraine
[#] Corresponding author: farkhad.aliev@uam.es

Ferromagnetic dots with magnetic vortex ground state [1] attract increasing interest in the last years due to the possibility of using the magnetic vortex state as a key element of a new type of high areal density storage media and importance of vortex spin excitations for operations of spin-polarized current driven spintronic devices as well as logic operation devices [2,3]. Implementation of the vortices in information storage assumes using both the vortex core magnetization direction (up or down) and vortex chirality (i.e., clockwise or anticlockwise rotation direction of the in-plane dot magnetization). Switching between these well defined states, which is expected to be done by either external field, electric current or their combined action, is accompanied by excitation of the spin waves which influence essentially the switching time. It is well established that a magnetic dot in the vortex ground state possesses two qualitatively different spin excitation modes: gyrotropic (the lowest excitation mode, typically below 1 GHz) [4] as well as high frequency (GHz) radial and azimuthal modes [5-7], which are labeled by the number of nodes ($n$) and ($m$) along the radial and the azimuthal directions in the circular dots, respectively.

Understanding of the high frequency spin wave excitation modes in ferromagnetic dots in the vortex state is of special fundamental and applied interest because knowledge of these modes allows controlling the magnetic vortex switching characteristic times and fields [1,2]. For instance, the azimuthal spin wave modes determine the vortex core distortion during the vortex switching in perpendicular magnetic field [8]. The spin eigenmode frequencies in micron and submicron size dots are determined mainly by magnetostatic interaction, and therefore depend on the dot aspect ratios. There were reported time domain measurements of excited by in-plane magnetic field the spin wave azimuthal modes with indices $n=0$, $m=\pm1$ for circular permalloy dots with small aspect ratio thickness/radius $\beta$ (below 0.06) [9]. It was found that the simulated eigenfrequencies and their splitting values were considerably higher than the experimental ones. Zhu et al. [10] reported on the lowest spin excitation frequencies for the dots with aspect ratio $\beta=0.087$ only. Recently Hoffman et al. [11] also measured the azimuthal eigenmode frequencies ($n=0$, $m=\pm1$) for dots with small aspect ratio (<0.03) and reported on the values which are substantially different from those measured by Park and Crowell [9].

Here we present room temperature broadband ferromagnetic resonance measurements of the zero magnetic field azimuthal and radial spin wave (SW) modes in square arrays of circular permalloy (Py) dots with moderate aspect ratio varying from 0.03 to 0.1. We observed two azimuthal frequency doublets with indices $n=0$, 1, $m=\pm1$ and one radial mode ($n=0$, $m=0$) for all the investigated samples. To the best of our knowledge this is the first systematic study of the high frequency azimuthal SW doublet with $n=1$.

Square arrays of polycrystalline Py dots with the thicknesses $L$=50, 25, 20, 15 nm and radii $R$=300, 500 nm were fabricated using electron beam (EB) lithography and lift-off techniques. Double layer spin coating resist over a Si wafer and highly directional EB evaporation allowed obtaining dots with sharp edges. The patterned area had dimensions from 2×2 to 10×10 mm$^2$. Details on the high frequency measurements of magnetization dynamics using a vector network analyzer (VNA) can be found in Refs. 12, 13. Although in previous time resolved scanning Kerr microscopy (TRKM) experiments [9-11] it was possible to probe locally smaller areas of the individual dots in time domain, the VNA-FMR technique provides more precise detection of the SW modes in the vortex state in frequency domain. In addition, the VNA-FMR provides the ability of averaging multiple spectra taken with relatively small excitation field as well as the ability to measure the response of dots in an array [14] that increases the signal to noise ratio in comparison with the TRKM experiments. We note that in our VNA-FMR setup, sketched in Fig.1a, in addition to the main azimuthal modes (having $m$ =±1) some radial ($n=0$, 1..., $m=0$) modes might be also excited [15] as the used samples are wider than the waveguide.

Figure 1b compares the experimentally observed SW modes #(1-5) and micromagnetic simulations [12] for the dot array with $L$=25 nm and $R$=518 nm ($\beta$=0.0483). Both the curves are normalized by the frequency of the lowest mode. The origin of a small (of about 10%) difference between the simulated and the detected frequencies (for example, the lowest simulated azimuthal mode is 5.62 GHz, whereas it is 5.05 GHz in the experiment) could be a weak dipolar coupling between the dots, the fact that simulations are done at T=0 K, and an error in saturation magnetization value taken in the simulations. Four experimentally detected modes (#1,2,4,5) correspond well to four simulated azmuthal modes ($n$=0,1; $m$=±1). The mode #3 situated between the first ($n$=0, $m$=±1) and second ($n$=1, $m$=±1) doublets of the split azimuthal modes is absent in the simulations. It is, most probably, a radially symmetric mode (this mode has a node in the dynamical magnetization profile in the middle of the dot with respect to the radial coordinate) excited by the small perpendicular field component created in the waveguide due to finite size of the central conductor (Fig. 1a). The field dependence of the modes #4, 5 is well described by simulations [12], which consider only in-plane rf driving field. This supports interpretation of the mode #3 as a radial mode and the modes #4, 5 as azimuthal modes for β=0.048, Fig. 1c shows modes #3, 4, 5 intensity plots of the measured SW spectra for magnetic field below the vortex nucleation field (β=0.04). The second doublet of the azimuthal modes ($n$=1, $m$=±1) indicated by green stars and blue circles shows reduced intensity in comparison with the first doublet due to existence of a node of the dynamic magnetization along the radial direction. We find that the lower azimuthal mode ($n$=1) frequency shows strong parabolic dependence, while the upper one is influenced weakly by the magnetic field. For the aspect ratios β=0.03 - 0.1 the frequency of the main radial mode (0, 0) is always close to the frequencies of the second azimuthal doublet ($n$=1, $m$=±1). More detailed analysis of the mode #3,4,5 intensities (the azimuthal doublet modes should have almost equal intensities) shows that the mode #3 is radial for the samples with β= 0.04, 0.048, 0.067, and 0.083, but the radial mode is high frequency mode #5 for the samples with β= 0.1. i.e., the dependence of the radial mode frequency $\omega_{0,0}(\beta)$ vs. β has a higher slope than the similar dependencies for the azimuthal modes $\omega_{1,\pm1}(\beta)$. There were just one mode above the first frequency doublet observed for β=0.03 and 0.05 ($L$=15 nm).

Influence of the interdot dipolar interaction was studied in dot arrays for static magnetization and it was shown that for the interdot separation $d$ (d=CC-2R where CC is the center to center distance between the dots) exceeding the dot radius the interaction can be neglected [16-18]. Influence of the dipolar interaction on the excited mode frequencies was evaluated using a set of dot arrays with the same aspect ratio and the largest thickness (radius $R$=500 nm and thickness $L$=50 nm, β =0.1) but with (CC) varied from 1200 nm to 2500 nm. Figure 1d shows the vortex state eigenfrequencies as a function of the ratio δ= d/R (the relative interdot separation). One clearly observes that the excited modes trend to increase their frequencies with increasing δ due to decreasing of (negative) interdot dipolar field that will be considered in detail elsewhere. Relatively small change of the eigenfrequencies seen for CC increasing from 1500 nm to 2500 nm allows neglecting dipolar interactions for the dot arrays with δ ≥ 1.5. Our analysis of the dependence of the excited spin eigenmodes on the dots aspect ratio was always done for the dots with δ ≥ 1.5.

The ($n$=1, $m$=±1) azimuthal modes were clearly detected only for samples with the dot thickness of 20, 25 and 50 nm. Figures 2 and 3 summarize our main experimental findings: (i) the experimental ($n$=0, 1; $m$=±1) azimuthal eigenmode frequencies and their splitting are in a good agreement with the calculated frequencies within a theoretical model described below; (ii) the second doublet ($n$=1, $m$=±1) frequencies and their splitting are observed for the first time. The splitting proves the conception of the moving vortex interaction with all the azimuthal SW ($n$=0, 1,...) having the same azimuthal symmetry ($m$=±1) as the vortex gyrotropic mode. Figure 2 demonstrates that for β< 0.05 the low frequency azimuthal modes ($n$=0, $m$=±1) follow predicted theoretically [5] and observed experimentally ~ $\beta^{1/2}$ dependence vs. β (the open triangles data in Fig. 2 are taken from Ref. 11), whereas for β exceeding 0.05 both the ($n$=0, $m$=±1) modes show downwards deviation from the linear dependence ~ $\beta^{1/2}$ mainly due to decreasing of $z$-component of the dynamic magnetostatic field. The correspondence of the sign of $m$ to lower/upper frequency in the azimuthal doublet depends on the vortex core polarization. We measured the dot arrays with different (random) core polarizations, so this sign has no direct physical sense, and used just to number the azimuthal modes in the spectra. We attribute small mismatch between present and previously measured for small β (<0.03) azimuthal mode frequencies (Fig. 2) to differences in the saturation magnetization of Permalloy dot arrays. We note also that the lowest azimuthal eigenmode frequencies ($n$=0, $m$=±1) measured for the single value of β= 0.087 by Zhu et al. [10] are also in reasonably good agreement with the present VNA-FMR data.

Figure 3 shows dependence of the measured and calculated frequency splitting of the first (Fig. 3a) and the second (Fig. 3b) azimuthal modes on the dot aspect ratio. While the splitting of the first azimuthal mode ($n=0$, $m=\pm 1$) shows nearly linear dependence on the dots aspect ratio almost in the all range of $\beta$ studied, the second azimuthal mode ($n=1$, $m=\pm 1$) splitting reveals non monotonous behaviour with clear maximum for $\beta$ around of 0.05. Such behaviour is due to change of the SW mode radial profiles increasing $\beta$.

For interpretation of the observed spin excitations we introduce the reduced magnetization $\mathbf{m}(\mathbf{r},t) = \mathbf{M}(\mathbf{r},t)/M_s$, $\mathbf{m}^2 = 1$, and assume that $\mathbf{m}$ does not depend on $z$-coordinate along the dot thickness. We apply the Landau-Lifshitz equation of motion $(M_s/\gamma)\partial \mathbf{m}/\partial t = \mathbf{m} \times (\partial w/\partial \mathbf{m})$, where $w$ is the energy density $w = -\mathbf{M} \cdot \mathbf{H}_m / 2$, $\mathbf{H}_m$ is the magnetostatic field. To find the vortex high-frequency modes, the linearized equations of motion of dynamic $\mathbf{m}$ over the vortex ground state $\mathbf{m}_0(\rho)$ are solved. They are linearized by the substitution $\mathbf{m}(\boldsymbol{\rho},t) = \mathbf{m}_0(\rho) + \boldsymbol{\mu}(\boldsymbol{\rho},t)$, where small $\boldsymbol{\mu} = (\mu_\rho, \mu_\varphi, \mu_z)$, $|\boldsymbol{\mu}| \ll 1$ corresponds to SW excitations, and $\rho, \varphi$ are the polar coordinates. The $\boldsymbol{\mu}$-components are used in the form of waves $\mu_\rho(\boldsymbol{\rho},t) = b(\rho)\sin(m\varphi - \omega t)$, $\mu_z(\boldsymbol{\rho},t) = a(\rho)\cos(m\varphi - \omega t)$ travelling in azimuthal direction, where $a$, $b$ are the SW amplitudes and $\mu_\varphi = 0$. In zero-bias field case only the eigenmodes with $m=\pm 1$ can be excited by uniform in-plane rf field. Neglecting the small dynamic exchange interaction we reduce the problem to eigenvalue problem for the integral magnetostatic operator [19]. The solution of this problem for a thin dot yields a discrete set of magnetostatic SW eigenfunctions $a_n(\rho)$, $b_n(\rho)$ and corresponding eigenfrequencies $\omega_{n,m}$, which are well above the gyrotropic mode eigenfrequency.

The azimuthal SW modes with $m = \pm 1$ have the same angular dependence as the vortex gyrotropic mode, and this leads to a considerable frequency splitting of the degenerated SW $m = \pm 1$ modes due to moving vortex core. Generalizing the results of Ref. 13 for the case of arbitrary radial index $n$ the azimuthal SW frequency ($m=+1/-1$) splitting is calculated as $\Delta\omega_{n,\pm 1} = \Im_n^2 (\omega_{n1}^2 - 2\omega_{nD}^2)/2N_n^2$, where $\Im_n = \int d\rho \rho m_0(\rho) b_n(\rho)$ is the overlapping integral of the vortex gyrotropic mode $m_0(\rho) = (1-\rho^2)/\rho$ [20] and the SW eigenmodes $b_n(\rho)$, $\omega_{n1}$ are the SW eigenfrequencies without vortex core ($|m| = 1$) similar to used in Ref. 20, and $\rho$ is in units of $R$.

$N_n = \int d\rho \rho b_n^2(\rho)$, $\omega_{nD}^2 = N_n^{-1} \int_0^\infty dk k^{-1} f(\beta k) I(k) I_n(k)$ describes the inter-mode dipolar coupling,

$I(k) = 2\int_0^1 d\rho \rho J_1(k\rho)$, $I_n(k) = \int_0^1 d\rho \rho b_n(\rho) \partial J_1(k\rho)/\partial \rho$, and $f(x) = 1 - (1 - \exp(-x))/x$. The azimuthal

modes profiles are shown in Fig. 1 of Ref. 20.

The SW magnetostatic frequencies are $\omega_{n1}^2 = \omega_M^2 F(L/R)$, where the function $F(\beta)$ depends only on $\beta$ and on the magnetization distribution of the $n$-eigenmode $\boldsymbol{\mu}_n(\boldsymbol{\rho})$. The function $F(\beta)$ to a good approximation is linear in $\beta = L/R$ and, therefore, we get the simple relations $\omega_{n1} \propto \beta^{1/2}$ and

$\Delta\omega_{n,\pm1} \propto \beta$ at $\beta \ll 1$. The splitting of $m = \pm1$ modes is a result of interaction of the high frequency

SW with the moving vortex core. The calculation of the azimuthal SW frequencies allows to explain quantitatively the experimental data presented in Fig. 3 for the mode indexes $n$=0, 1 and $m$=+1/-1.

The frequency splitting for $n$=1 is approximately twice smaller than for $n$=0 azimuthal modes due to decrease of $\Im_n$ with $n$ increasing. The radial ($m$=0) mode frequencies can be described by the equation $\omega_{n0}^2 = \omega_M^2 f(\beta\alpha_{1n})$, where $\alpha_{1n}$ is the $n$-th root of the equation $J_1(x) = 0$ [19]. The value of

$\omega_M = 4\pi\gamma M_s$ =29.7 GHz was used to calculate the frequencies, which is an average of $\omega_M$ extracted

from fitting the FMR spectra of the saturated dot arrays with different $\beta$.

*In conclusion*, the main SW eigenmodes, namely the first, second azimuthal and first radial

SW modes, have been detected and studied in Py magnetic dot arrays as function of the dot aspect ratio in the range 0.03 - 0.1. The observed splitting of the degenerated azimuthal doublets ($n$=0,1,

$m$=±1) is in good agreement with the model of the magnetostatic modes in the vortex state dots that

takes into account dynamical origin of the azimuthal modes frequency splitting. In contrast to the first doublet, the azimuthal mode frequency splitting of the second doublet shows a maximum as function of the dot aspect ratio. Understanding of the high frequency spin dynamics in the vortex state magnetic dots with moderate aspect ratios can be used for manipulation of the future generation of magnetic memories and spin torque nano-oscillators.

The work in Madrid (A.A.A., J.F.S., and F.G.A.) was supported by Spanish MICINN (MAT2009-10139), Consolider 'Molecular Nanoscience' (CSD2007-00010) and CAM (S-505/MAT0194). K.G. acknowledges support by the Ikerbasque Science Foundation. G.N.K. acknowledges support from Portuguese FCT through "Ciencia 2007" program.

Figure 1. (a) Sketch of the VNA-FMR setup. The inset shows tilted SEM image of the dots. (b) Dynamic magnetic response simulated [12] and measured for Py dots with aspect ratio $\beta= 0.048$ normalized by the frequencies of the first mode. Vertical dotted lines and symbols show relation between the simulated and measured eigenmodes #1, 2, 4, 5. (c) Intensity plots of the measured spin wave spectra for magnetic field below the vortex nucleation field. The magnetic field is applied along [11] dot array lattice direction and is perpendicular to RF field ($h_{rf}$). (d) Dependence of the lowest excited eigenmode frequencies on the relative interdot separation $\delta = (CC\text{-}2R)/R$ for Py dots (the dot radius is $R$=500 nm and the thickness is $L$= 50 nm, aspect ratio $L/R$= 0.1).

Figure 2. The measured spin wave mode frequencies vs. square root of the dot aspect ratio $\beta$. All experimental data are taken for maximum interdot separation in the Py dot arrays. The dashed line represents theoretical values described in the text. Open triangles show the lowest azimuthal mode frequencies [11] measured for very thin dots with $\beta < 0.03$.

Figure 3. (a) Measured splitting (solid circles) of the first azimuthal mode ($n$=0, $m$=±1, red and blue triangles in Fig. 2) as function of the dot aspect ratio $\beta$. The dashed line represents theoretical values described in the text. Black star shows the lowest azimuthal mode splitting measured by Zhu et al. [10]. (b) The second azimuthal doublet frequency spitting ($n$=1, $m$=±1, dark blue circles and green stars in Fig. 2) vs. dot aspect ratio $\beta$.

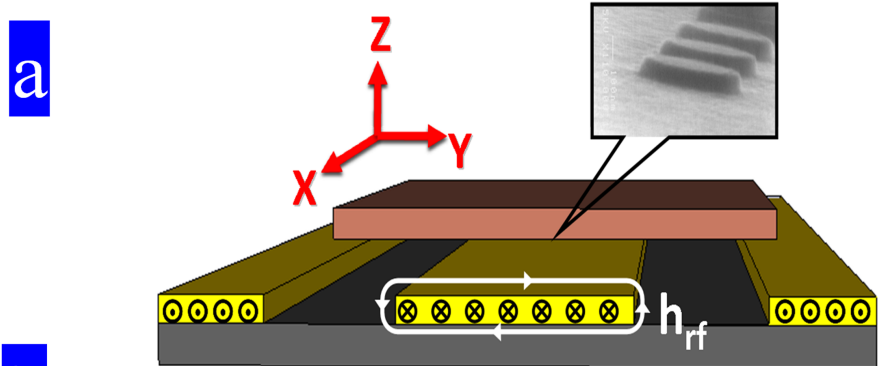

a

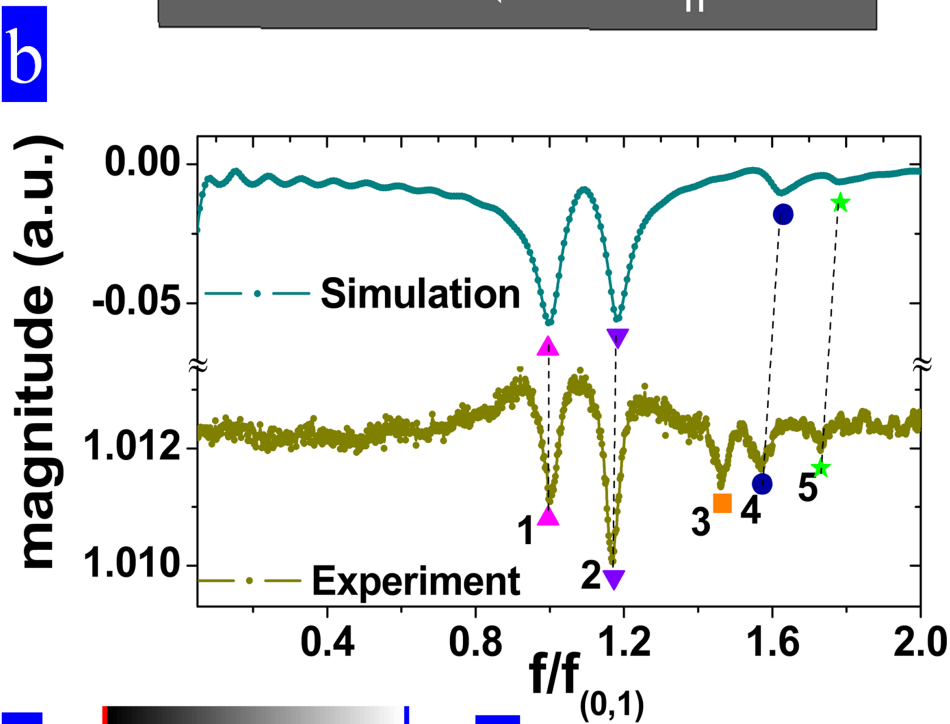

b

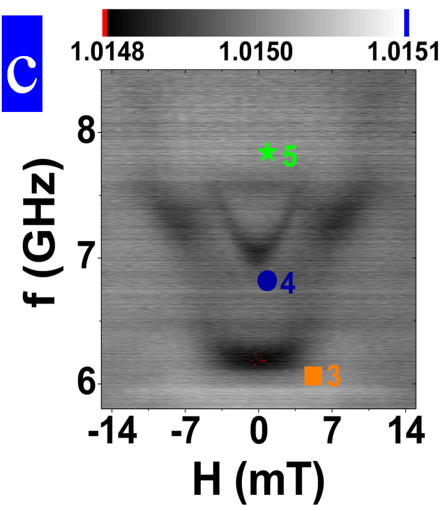

c

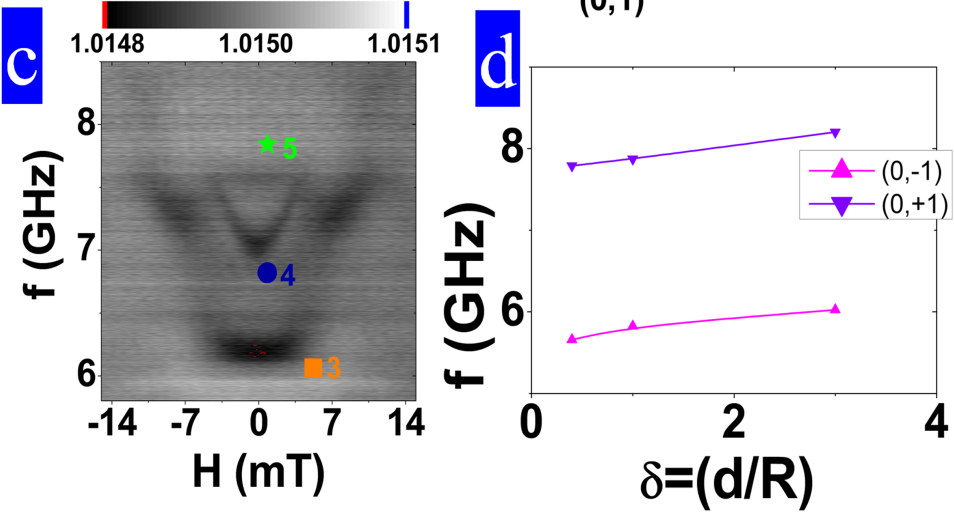

d

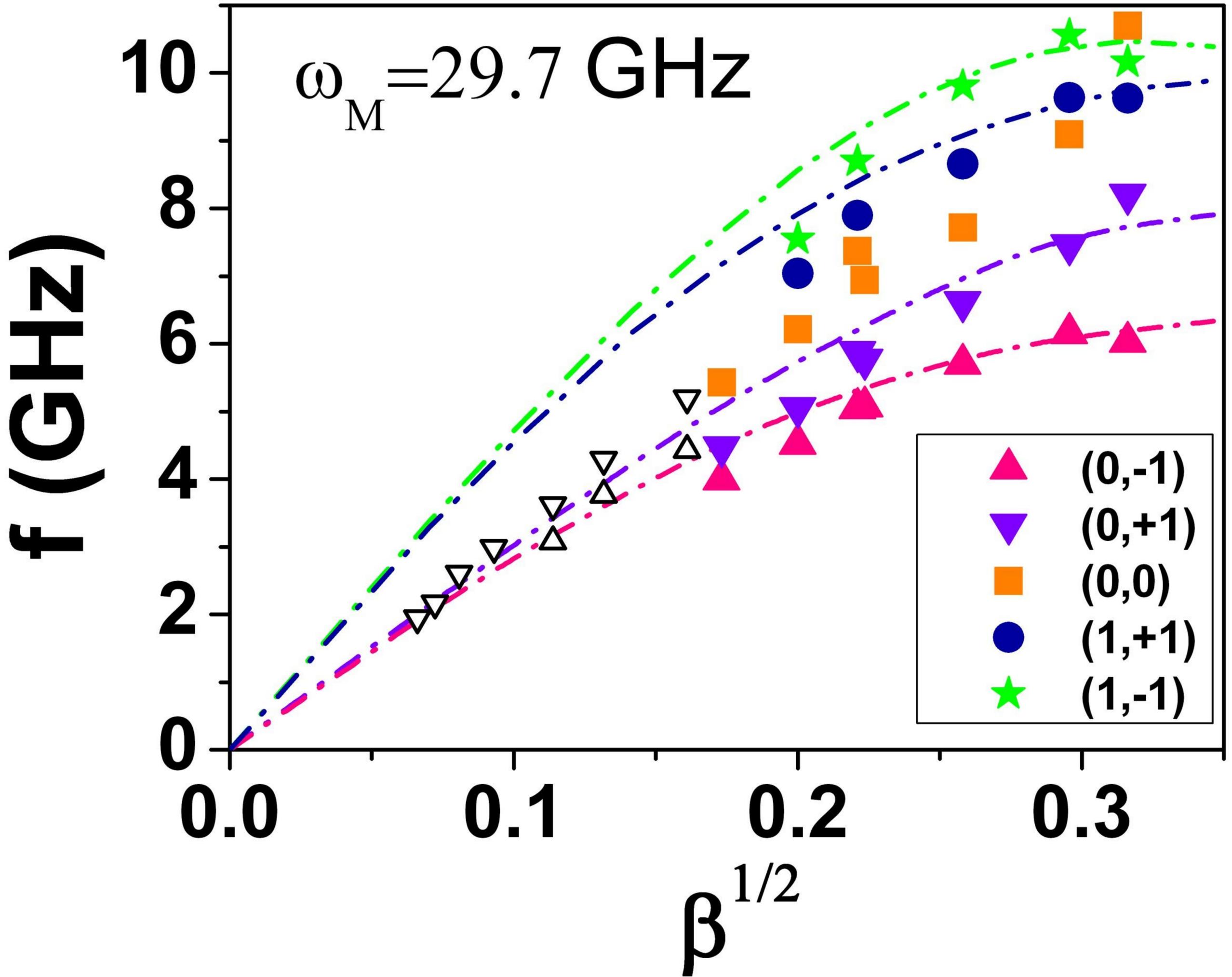

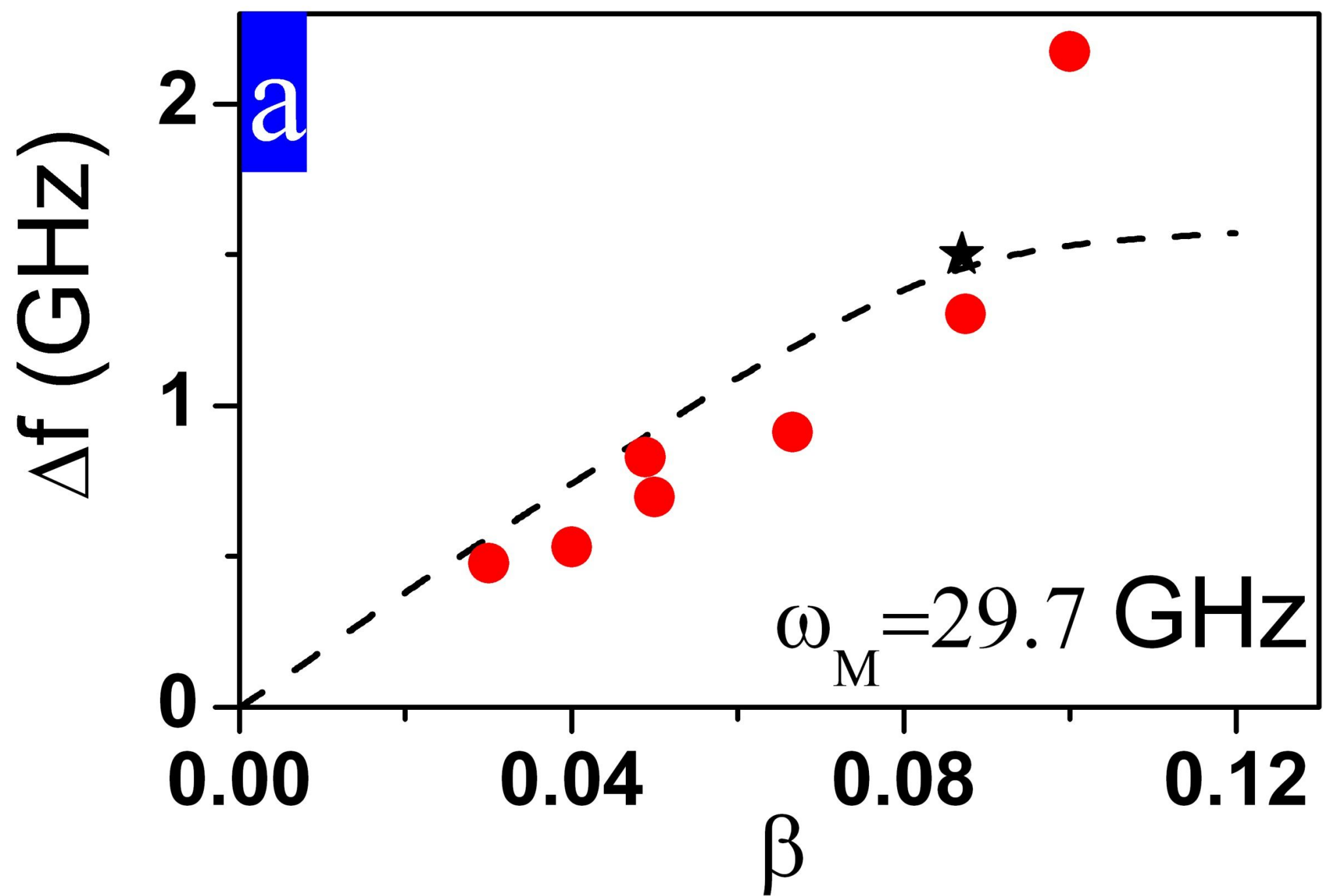

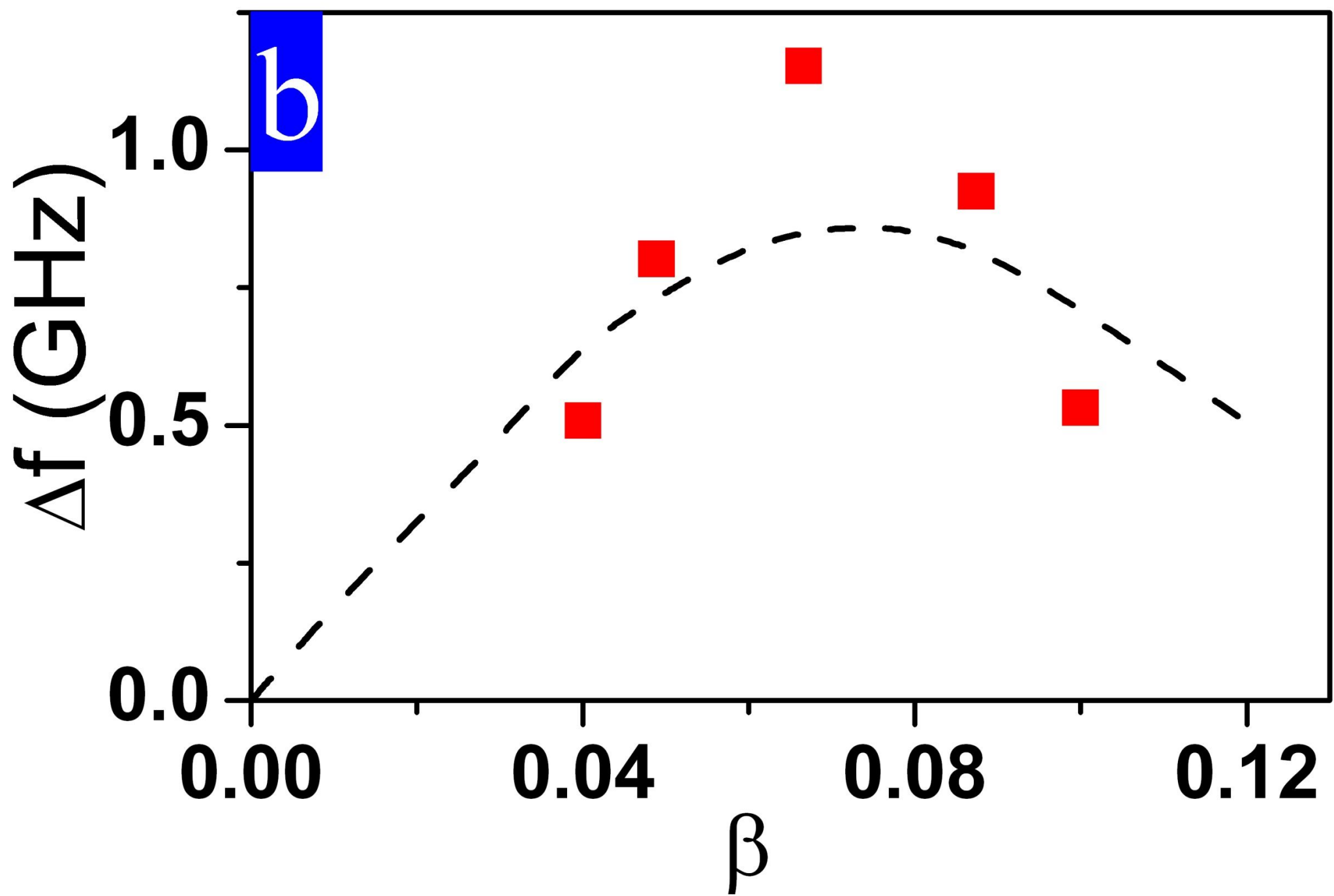